\documentclass[superscriptaddress,pre,aps,twocolumn,showpacs,floatfix]{revtex4}

\usepackage[dvips]{graphicx}
\usepackage{amsmath}
\usepackage{amssymb}
\usepackage[nice]{nicefrac}
\usepackage{color}
\usepackage{amsmath}
\usepackage{grffile}
\usepackage{times}
\usepackage{grffile}

\newcommand{\sgn}{\mbox{sign}}

\renewcommand{\=}{\stackrel{d}{=}}

\begin{document}
\title{Resonant activation in 2D and 3D systems driven by multi-variate L\'evy noises}

\author{Krzysztof Szczepaniec}
\email{kszczepaniec@th.if.uj.edu.pl}
\affiliation{Marian Smoluchowski Institute of Physics, and Mark Kac Center for Complex Systems Research, Jagiellonian University, ul. Reymonta 4, 30--059 Krak\'ow, Poland }

\author{Bart{\l}omiej Dybiec}
\email{bartek@th.if.uj.edu.pl}
\affiliation{Marian Smoluchowski Institute of Physics, and Mark Kac Center for Complex Systems Research, Jagiellonian University, ul. Reymonta 4, 30--059 Krak\'ow, Poland }

\date{\today}

\begin{abstract}
Resonant activation is one of classical effects demonstrating constructive role of noise.
In resonant activation cooperative action of barrier modulation process and noise lead to the optimal escape kinetics as measured by the mean first passage time.
Resonant activation has been observed in versatilities of systems for various types of barrier modulation processes and noise types.
Here, we show that resonant activation is also observed in 2D and 3D systems driven by bi-variate and tri-variate $\alpha$-stable noises.
Strength of resonant activation is sensitive to the exact value of the noise parameters. 
In particular, the decrease in the stability index $\alpha$ results in the disappearance of the resonant activation.

\end{abstract}

\pacs{
 05.40.Fb, 
 05.10.Gg, 
 02.50.-r, 
 02.50.Ey, 
 }
\maketitle

\section{Introduction\label{sec:introduction}}

Typically noise is considered as an unwanted disturbance to data. In many cases, sophisticated filtering methods have been developed in order to purify recorded signals. Paradoxically, presence of noise can be also beneficial. Last decades have witnessed growing interest in the so called noise induced phenomena, i.e. various types of effects which occur because of the presence of noise. In complex realms, not fully known interactions can be approximated by noise. Consequently, noise can be used to simplify effective description of dynamical systems when the detailed character of interactions is unknown or too complicated for exact methods.

Among noise induced effects stochastic resonance \cite{gammaitoni1998,anishchenko1999} and resonant activation \cite{doering1992,boguna1998} are considered as seminal examples of effects demonstrating constructive role of fluctuations, which put our understanding of the role of noise into new direction. The growing number of studies examining various systems perturbed by noise demonstrated that high efficiency of dynamical systems can originate due to fluctuations.
In stochastic resonance weak periodic signals are amplified by noise. The optimal noise level results in maximal signal amplification, as measured by the signal to noise ratio or spectral power amplification \cite{gammaitoni1998}. Consequently, weak sub-threshold signals become detectable. Resonant activation is an effect of the optimal noise assisted escape over fluctuating potential barrier. Combined action of noise and barrier modulation process results in the shortest (minimal) escape time --- mean first passage time.

The most common, as well as the simplest, approximation of complex interactions is Gaussian white noise, which is used as an archetypal process modeling complex interactions of a test particle with its environment.
Gaussian white noise approximation works perfectly well when interactions are independent and bounded.
Both these assumptions, i.e. boundeness and whiteness, can be violated resulting in more general non-Markovian and non-Gaussian processes. Such extensions allow for description of more complex realms.
In this context, the special role is played by white $\alpha$-stable noises which are capable to describe out-of-equilibrium systems displaying heavy-tailed fluctuations \cite{samorodnitsky1994,metzler2000,chechkin2006,klages2008,dubkov2008,metzler2007} and outliers.
Growing number of experimental observations \cite{shlesinger1995,nielsen2001,ditlevsen1999b,mantegna2000,solomon1993,chechkin2002b,boldyrev2003} demonstrate that heavy-tailed fluctuation are ubiquities in the nature. Experimental observations have advances theoretical models based on $\alpha$-stable noises stimulating research on stochastic resonance \cite{kosko2001,applebaum2009,dybiec2009e,dybiec2009} and resonant activation \cite{dybiec2007b,dybiec2004,dybiec2007,dybiec2009,munakata1985,bag2003,hanggi1994,iwaniszewski2003,novotny2000,majee2005} in systems driven by more general than Gaussian white noise or even in non-Markovian realms \cite{szczepaniec2014quantifying}.

Within the current manuscript the resonant activation phenomenon driven by bi-variate and tri-variate $\alpha$-stable noises is studied. Such an extension not only allow to study resonant activation in non-equilibrium realms but also explore role of increasing spatial dimensionality.
The studied model is presented in Section~\ref{sec:model}.
Section~\ref{sec:results} discusses obtained numerical results.
The paper is closed with concluding remarks (Section~\ref{sec:summary}).

\section{Model \label{sec:model}}

$\alpha$-stable densities are limiting distribution for sums of independent identically distributed random variables, even in situations when variables are characterized by the infinite second moment. Properties of $\alpha$-stable variables: invariance under convolution, power-law asymptotics and self-similarity make them especially suited for modeling of out-of-equilibrium situations when heavy tailed fluctuations are observed.

The random variable $X$ is said to be stable, if any linear combination of two independent copies of the variable is distributed according to the same distribution up to rescaling and shift, i.e.
\begin{equation}
 A{X}^{(1)}+B{X}^{(2)} \= C{X} + {D},
 \label{eq:definition}
\end{equation}
where $\=$ denotes equality in distributions.
The random variable $X$ is called strictly stable if $D=0$.
Finally, a random variable is $\alpha$-stable if Eq.~(\ref{eq:definition}) holds with $C=(A^\alpha+B^\alpha)^{1/\alpha}$ where $0 < \alpha \leqslant 2$.
The characteristic function of $\alpha$-stable random variables can be determined by the defining Eq.~(\ref{eq:definition}).
The Eq.~(\ref{eq:definition}) can be straight forward extended into higher dimensions in such a case $\boldsymbol{X}=(X_1,\dots,X_d)$ represents a random vector in $\mathbb{R}^d$.

The main scope of the current article is to study resonant activation in 2D and 3D systems driven by bi-variate and tri-variate $\alpha$-stable noises.
First, the archetypal model of resonant activation is presented (Sec.~\ref{sec:1d}).
Next, the RA model is extended into 2D (Sec.~\ref{sec:2d}).
In addition, the basic information about multi-variate $\alpha$-stable noises is included.

\subsection{Resonant activation in 1D systems\label{sec:1d}}

Typically resonant activation has been studied in 1D realms or in 2D systems which can be reduced to 1D models with entropic barriers
\cite{burada2008,burada2009entropic,mondal2010entropic}.
The motion of a particle is confined to a finite interval, e.g. $[0,1]$.
A particle starts its motion at the reflecting boundary, i.e. $x(0)=0$.
At the other end of the interval there is an absorbing boundary, i.e. every time a particle reaches the boundary or passes over the boundary it is immediately removed from the system.
The particle motion is described by the Langevin equation
\begin{equation}
 \frac{dx}{dt}=-V'(x,t)+\sigma\xi(t),
 \label{eq:langevin}
\end{equation}
where $-V'(x,t)$ is a time dependent force acting on a test particle and $\xi(t)$ is Gaussian white noises ($\langle \xi(t) \rangle=0$ and $\langle \xi(t) \xi(s) \rangle=\delta(t-s)$).
As in the seminal Doering Gadoua paper \cite{doering1992}, $V(x,t)$ is a time dependent linear potential switching dichotomously between two configurations $V_\pm(x)$ characterized by two distinct heights $H_\pm$, i.e. $V_\pm(x)=H_\pm x$. Initially, a test particle is located at the origin and the configuration of the potential is set to $V_+(x,t)$ or $V_-(x,t)$ with equal probabilities. It is assumed that the dichotomous process is Markovian and has the same switching rates. More precisely, it takes two values $a_\pm$ and it is described by a single parameter $\gamma$ which is the rate of the potential switching
\begin{equation}
 \begin{array}{rcccl}
 & & a_+ & & \\
 \gamma & \uparrow & & \downarrow & \gamma\\
 & & a_- & & \\
 \end{array}.
\end{equation}
The dichotomous process takes two possible values only and stays constant for the exponentially distributed time.
The autocorrelation of the dichotomous process is $\frac{1}{4}(a_+-a_-)^2\exp\left[ -2\gamma t \right]$, see \cite{horsthemke1984,gardiner2009}.

The first possible generalization of the model described by Eq.~(\ref{eq:langevin}) is to replace the Gaussian white noise with the more general white $\alpha$-stable L\'evy type noise $\zeta(t)$
\begin{equation}
 \frac{dx}{dt}=-V'(x,t)+\sigma\zeta(t),
 \label{eq:langevin-frac}
\end{equation}
or
\begin{equation}
 dx=-V'(x,t)dt+\sigma d L_{\alpha,0}(t).
 \label{eq:langevin-frac-incr}
\end{equation}
Increments of a symmetric $\alpha$-stable motion $L_{\alpha,0}(t)$ are independent and distributed according to a symmetric $\alpha$-stable density with the characteristic function $\phi(k)= \mathbb{E} \left[ e^{ikX} \right]$ given by \cite{samorodnitsky1994,janicki1994}
\begin{equation}
 \phi(k)= 
 \exp\left[ -\sigma^\alpha |k|^\alpha \left( 1 - i\beta\sgn k \tan\frac{\pi\alpha}{2} \right) +i\mu k \right] 
 \label{eq:characteristic1d}
\end{equation}
for $\alpha\neq 1$ and
\begin{equation}
 \phi(k)= 
 \exp\left[ -\sigma |k| \left( 1 + i\beta\frac{2}{\pi}\sgn k \ln |k| \right) + i\mu k \right] 
 \label{eq:characteristic1d-b}
\end{equation}
for $\alpha=1$.
$\alpha \in (0,2]$ is the stability index, $\beta \in [-1,1]$ is the asymmetry (skewness) parameter, $\sigma > 0$ is the scale parameter and finally $\mu \in \mathbb{R}$ is the location parameter. The closed formulas for $\alpha$-stable densities are known only in a limited number of cases: $\alpha=2$ -- normal distribution, $\alpha=1$ with $\beta=0$ -- Cauchy distribution and $\alpha=\nicefrac{1}{2}$ with $\beta=1$ -- L\'evy-Smirnoff distribution.
In general, symmetric $\alpha$-stable densities with $\alpha<2$ have the power-law asymptotics of $|x|^{-(\alpha+1)}$ type.

For $\alpha<2$, the Langevin equation~(\ref{eq:langevin-frac}) is associated with the set of following (space) fractional time dependent diffusion equations \cite{yanovsky2000,schertzer2001,metzler1999,jespersen1999}:
\begin{eqnarray}
 \frac{\partial p_\pm(x,t)}{\partial t} & = & \left[ \frac{\partial}{\partial x} V'_\pm(x) + \sigma^\alpha \frac{\partial^\alpha}{\partial |x|^\alpha} \right]{} p_\pm(x,t) \\ \nonumber
 & & \mp\; \gamma p_\pm(x,t) \pm \gamma p_\mp(x,t),
\label{eq:ffpe}
\end{eqnarray}
where $p_\pm(x,t)$ are probabilities to find a particle in the vicinity of $x$ if the potential is in the $V_\pm(x)$ configuration.
In Eq.~(\ref{eq:ffpe}), $\frac{\partial^{\alpha}}{\partial |x|^{\alpha}}$ denotes the Riesz-Weil fractional space derivative defined by the Fourier transform
\begin{equation}
{\cal{F}}\left[\frac{\partial^{\alpha}}{\partial |x|^{\alpha}} f(x)\right]=-|k|^{\alpha} {\cal{F}}\left[ f(x) \right].
\end{equation}
For $\alpha=2$, the standard diffusion equation is recovered, i.e. $\frac{\partial^\alpha}{\partial |x|^\alpha} \to \frac{\partial^2}{\partial x^2}$.
Boundaries impose additional constraints on the probability densities $p_\pm(x,t)$ due to reflection at $x=0$ and absorption at $x=1$. For $\alpha=2$, these boundary conditions are local \cite{gardiner2009}, otherwise they are non-local \cite{dybiec2006,zoia2007}.
The mean first passage time can be calculated as \cite{hanggi1990,doering1992}
\begin{equation}
 \tau = \int_0^\infty dt \int_0^1\left[ p_+(x,t)+p_-(x,t) \right]dx.
\end{equation}

\subsection{Resonant activation in 2D systems\label{sec:2d}}

The resonant activation setup can be extended to higher dimensions in the straight forward manner
\begin{equation}
 \frac{d\boldsymbol{r}}{dt}=-\nabla V(\boldsymbol{r},t)+\sigma \boldsymbol{\zeta}(t),
 \label{eq:langevin-2d}
\end{equation}
where 
$-\nabla V(x,t)$ is a time dependent force acting on a test particle and $\boldsymbol{\zeta}$ is a multi-variate white $\alpha$-stable noise. Alternatively, it is possible to use the incremental notation
\begin{equation}
d\boldsymbol{r} =-\nabla V(\boldsymbol{r},t)dt+\sigma d\boldsymbol{L}_{\alpha}(t).
 \label{eq:langevin-2d-incr}
\end{equation}
Increments of the 2D $\alpha$-stable motion $\boldsymbol{L}_{\alpha}(t)$ are independent and distributed according to the bi-variate $\alpha$-stable density which is a special $d=2$ example of multi-variate $\alpha$-stable densities. As in 1D case a white $\alpha$-stable noise $\boldsymbol{\zeta}_{\alpha}(t)$ is a formal time derivative of the $\alpha$-stable motion $\boldsymbol{L}_{\alpha}(t)$.
The characteristic function $\phi(\boldsymbol{k}) = \mathbb{E} \left[ e^{i \langle \boldsymbol{k},\boldsymbol{X}\rangle} \right]$ of the $\alpha$-stable vector $\boldsymbol{X}=(X_1,\dots,X_d)$ in $\mathbb{R}^d$ is given by \cite{samorodnitsky1994}

\begin{widetext}
\begin{equation}
 \phi(\boldsymbol{k}) =
 \left\{
 \begin{array}{lcl}
 \exp\left\{ -\int_{S_d} |\langle \boldsymbol{k},\boldsymbol{s} \rangle|^\alpha \left[ 1 -i\sgn(\langle \boldsymbol{k},\boldsymbol{s} \rangle)\tan\frac{\pi\alpha}{2} \right]\Gamma(d\boldsymbol{s}) +i \langle \boldsymbol{k},\boldsymbol{\mu}^0 \rangle \right\} & \mbox{for} & \alpha\neq 1,\\
 \exp\left\{ -\int_{S_d} |\langle \boldsymbol{k},\boldsymbol{s} \rangle|^\alpha \left[ 1 +i\frac{2}{\pi}\sgn(\langle \boldsymbol{k},\boldsymbol{s} \rangle)\ln(\langle \boldsymbol{k},\boldsymbol{s} \rangle) \right]\Gamma(d\boldsymbol{s}) +i \langle \boldsymbol{k},\boldsymbol{\mu}^0 \rangle \right\} & \mbox{for} & \alpha = 1,
 \end{array}
 \right.
 \label{eq:characteristicdd}
\end{equation}

\end{widetext}
where $\langle \boldsymbol{k} , \boldsymbol{s} \rangle$ represents the scalar product, $\Gamma(\cdot)$ stands for the (finite) spectral measure on the unit sphere $S_d$ of $\mathbb{R}^d$ and $\boldsymbol{\mu}^0$ is a vector in $\mathbb{R}^d$, see \cite{samorodnitsky1994}.

The spectral measure $\Gamma(\cdot)$ replaces asymmetry ($\beta$) and scale ($\sigma$) parameters which characterize 1D $\alpha$-stable densities, see Eqs.~(\ref{eq:characteristic1d}) ans (\ref{eq:characteristic1d-b}). Consequently, the spectral measure $\Gamma(\cdot)$ includes information about symmetry and width (scale) of multi-variate $\alpha$-stable densities, see \cite{samorodnitsky1994}.
Multi-variate $\alpha$-stable density is symmetric if the spectral measure is symmetric.
Usually, components of multi-variate $\alpha$-stable variables are dependent \cite{samorodnitsky1994}. 
Here, we will use bi-variate and tri-variate $\alpha$-stable variables with uniform spectral measures only, i.e. continuous spectral measures which are constant on the unit sphere $S_2$ or $S_3$.


Equation~(\ref{eq:langevin-2d}) can be integrated numerically using the stochastic Euler method \cite{higham2001,janicki1994,janicki1994b}.
Random numbers distributed according to multi-variate $\alpha$-stable densities can be generated using methods described in \cite{nolan1998b,samorodnitsky1994}. Nevertheless, trajectories of $\alpha$-stable motions can be also approximated by other methods \cite{teuerle2009,teuerle2012}. In general, the shape of $\alpha$-stable motions is determined by properties of the multi variate $\alpha$-stable densities. 
 
For uniform spectral measures, analogously like in 1D, the Langevin equation~(\ref{eq:langevin-2d}) can be associated with the following (space) fractional Smoluchowski-Fokker-Planck equation
 \begin{eqnarray}
 \frac{\partial p_\pm(\boldsymbol{r},t)}{\partial t} & = & 
\nabla \cdot \left[ \nabla V_\pm(\boldsymbol{r}) p_\pm(\boldsymbol{r},t) \right] 
- \sigma^\alpha (-\Delta)^{\alpha/2} p_\pm(\boldsymbol{r},t) \nonumber \\
& & 
\mp \; \gamma p_\pm(\boldsymbol{r},t) \pm \gamma p_\mp(\boldsymbol{r},t),
 \label{eq:ffpe2d}
 \end{eqnarray}
where $-(-\Delta)^{\alpha/2}$ is the fractional Riesz-Weil derivative (laplacian) defined by its Fourier transform \cite{samko1993} 
 \begin{equation}
 \mathcal{F}\left[ -(-\Delta)^{\alpha/2} p(\boldsymbol{r},t) \right] = -|\boldsymbol{k}|^\alpha \mathcal{F}\left[ p(\boldsymbol{r},t) \right].
 \label{eq:weil2d}
 \end{equation}
The drift term $\nabla \cdot \left[ \nabla V_\pm(\boldsymbol{r}) p_\pm(\boldsymbol{r},t) \right]$ originates due to the deterministic force $\boldsymbol{F}_\pm(\boldsymbol{r})=-\nabla V_\pm(\boldsymbol{r})$ acting on a test particle.
In Eq.~(\ref{eq:ffpe2d}) the drift term has the standard form, but the diffusive term depends on the noise type.
 For the bi-variate $\alpha$-stable noise with the uniform spectral measure the diffusion term is given by the fractional laplacian $-(-\Delta)^{\alpha/2}$, see \cite{samko1993}.
 In general, exact shape of the diffusion term is determined by the spectral measure $\Gamma(\cdot)$, see \cite{samko1993}.

The 2D resonant activation is studied for the escape from the disk with the absorbing edge. Initially a test particle is located in the center of the disk. The potential dichotomously switches between two configurations $V_\pm(\boldsymbol{r})=V_\pm(x,y)=H_\pm \sqrt{x^2+y^2}$. Thus, the force acting on a test particle is
\begin{equation}
\boldsymbol{F}_\pm(x,y)=-\nabla V_\pm(x,y)=-H_\pm\left[ \frac{x}{\sqrt{x^2+y^2}},\frac{y}{\sqrt{x^2+y^2}}\right]. 
\end{equation}
Consequently, the force is a 2D analog of a 1D linear force.
Due to non-local boundary conditions, the mean first passage time is calculated trajectory-wise using Monte Carlo methods.
The first passage time is defined as
\begin{equation}
 \tau = \min\{t>0 \;\;:\;\; \boldsymbol{r}(0)=\boldsymbol{0} \mbox{ and } |\boldsymbol{r}(t)| \geqslant 1 \},
 \label{eq:mfpt-trajectory}
\end{equation}
i.e. it is the earliest time when a particle leaves the prescribed domain of motion, i.e. disk of radius $R=1$. 
The particle starts its motion at $\boldsymbol{r}=\boldsymbol{0}$ and the absorbing boundary is located at $|\boldsymbol{r}|=1$.
The mean first passage time is the average first passage time $\langle \tau \rangle$.

\section{Results \label{sec:results}}

\begin{figure*}[!ht]
\begin{tabular}{ccc}

\includegraphics[angle=0,width=0.95\textwidth]{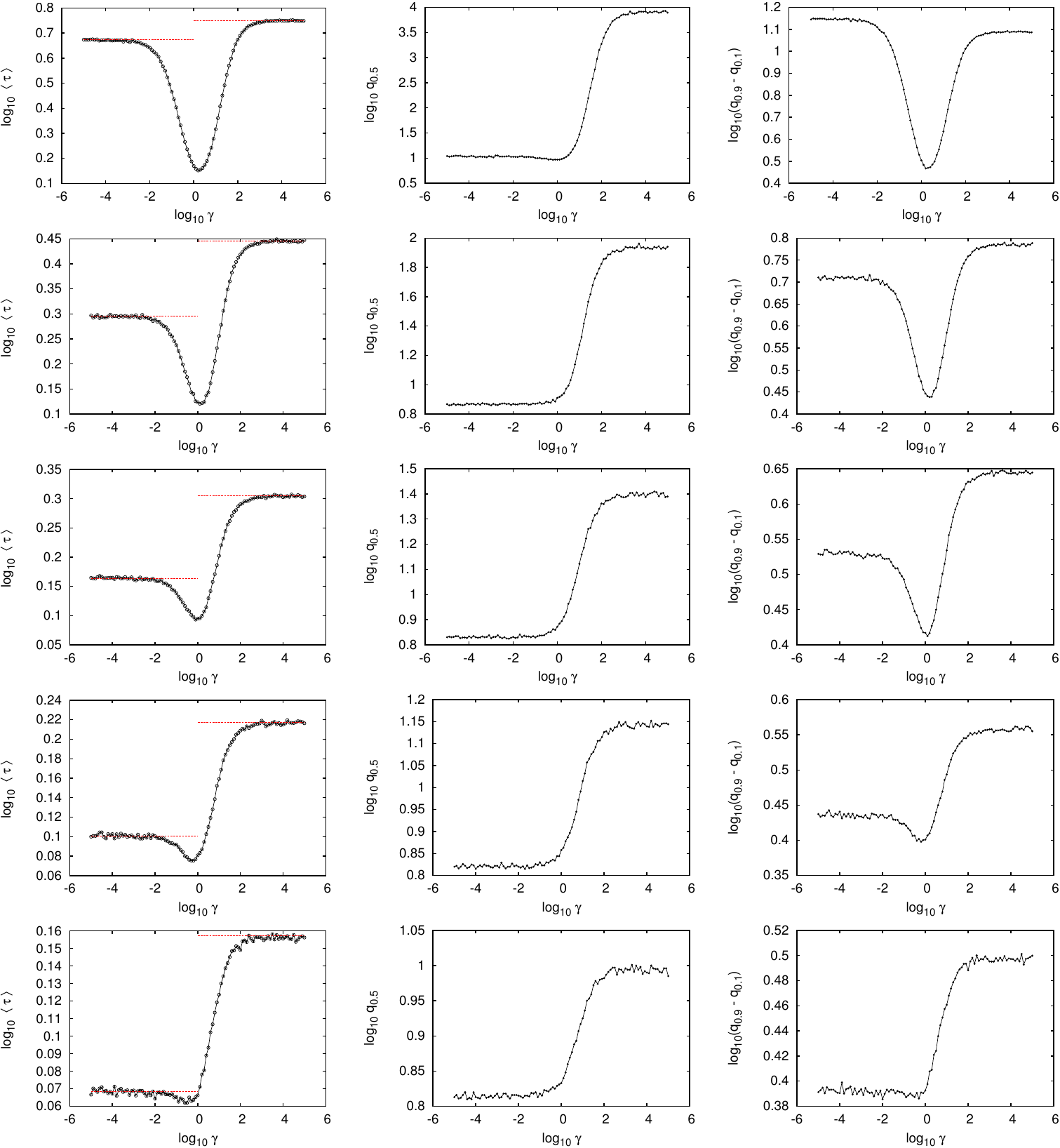} \\

 \end{tabular}
 \caption{Performance measures of the 2D resonant activation: mean first passage time $\langle \tau \rangle$ (left column), median of the first passage time distribution $q_{0.5}$ (middle column) and inter-quantile width of the first passage time distribution $q_{0.9}-q_{0.1}$ (right column). Various rows correspond to decreasing values of the stability index, $\alpha$, $\alpha \in \{1.9,1,7,1.5,1.3,1.1\}$ (from top to bottom). The potential dichotomously fluctuates between $V_\pm(\boldsymbol{r})=H_\pm r=H_\pm\sqrt{x^2+y^2}$ configurations with $H_+=8$ and $H_-=0$. The system is driven by the bi-variate $\alpha$ stable noises.
 Red dashed lines in the left column present low switching and high switching rate asymptotics.
 Error bars in the left panel represent standard deviation of the mean.
}
\label{fig:a19-a11}
\end{figure*}

Resonant activation is the effect of the optimal noise assisted escape kinetics over a fluctuating potential barrier.
Its presence is manifested by the minimal value of the mean first passage time, which is typically used to measure efficiency of the escape kinetics.
More precisely, for a fixed value of the noise parameters it is possible to optimize escape kinetics by adjusting the rate of the barrier modulation process.
In the situation, when potential switches dichotomously between two distinct configurations $V_\pm(\boldsymbol{r})$ the optimization of the barrier modulation process is performed by adjusting the rate $\gamma$ characterizing the switching (transition) rate.
Examination of the mean first passage time reveals that the resonant activation phenomenon is observed for the escape from 2D domains driven by bi-variate $\alpha$-stable noises.

\begin{figure}[!ht]
\begin{tabular}{cc}

\includegraphics[angle=0,width=0.95\columnwidth]{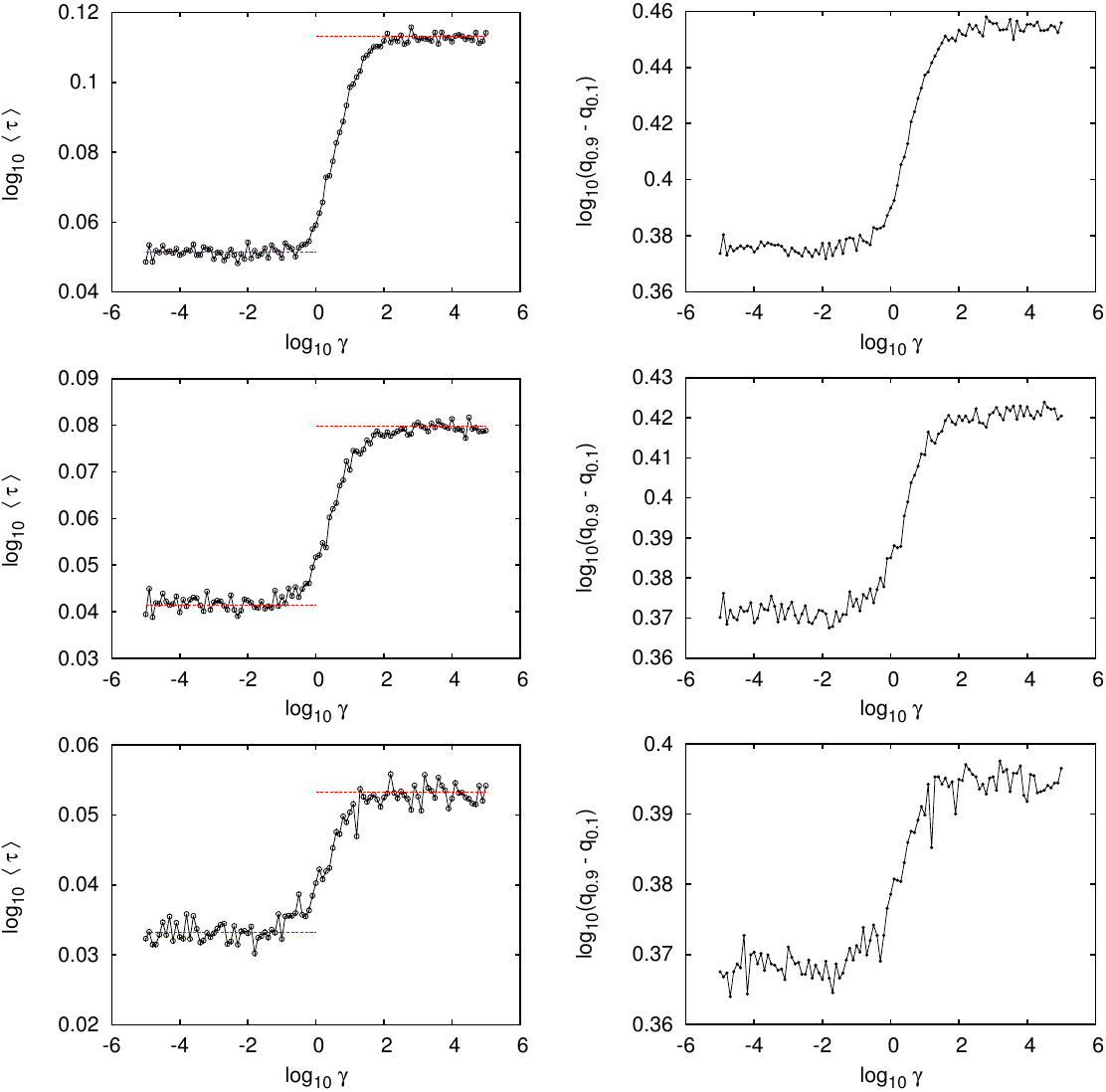} \\

 \end{tabular}
 \caption{Performance measures of the 2D resonant activation: mean first passage time $\langle \tau \rangle$ (left column) and inter-quantile width of the first passage time distribution $q_{0.9}-q_{0.1}$ (right column). Various rows correspond to decreasing values of the stability index, $\alpha$, $\alpha \in \{0.9,0,7,0.5\}$ (from top to bottom).
 The potential dichotomously fluctuates between $V_\pm(\boldsymbol{r})=H_\pm r=H_\pm\sqrt{x^2+y^2}$ configurations with $H_+=8$ and $H_-=0$. The system is driven by the bi-variate $\alpha$ stable noises.
 Red dashed lines in the left column present low switching and high switching rate asymptotics.
 Error bars in the left panel represent standard deviation of the mean.
}
\label{fig:a09-a07}
\end{figure}

Figure~\ref{fig:a19-a11} presents the mean first passage time (left panel), median of the first passage time density $q_{0.5}$ (middle panel) and the inter-quantile width $q_{0.9}-q_{0.1}$ of the first passage time density (right panel). Various rows correspond to decreasing values of the stability index $\alpha$ ($\alpha\in\{1.9,1.7,1.5,1.3,1.1\}$) (from top to bottom). The potential is dichotomously switching between two configurations $V_\pm(\boldsymbol{r})=V_\pm(x,y)=H_\pm \sqrt{x^2+y^2}$ with $H_+=8$ and $H_-=0$.
Examination of the mean first passage time demonstrates that the resonant activation is presented for all values of the stability index $\alpha>1$. With the decreasing value of the stability index $\alpha$ the strength of resonant activation weakens. Finally, for $\alpha<1$ and studied potentials $V_\pm(\boldsymbol{r})$, the effect of resonant activation disappears completely, see Fig.~\ref{fig:a09-a07}.

Resonant activation is the property of the system at hand but its presence affects the shape of the first passage time density.
Therefore, resonant activation can be quantified by some of measures which characterize width or location of the first passage time density \cite{szczepaniec2014quantifying}.
Changes in the efficiency of the escape kinetics affects characteristics of the first passage time density. Consequently, not only the mean value (mean first passage time) can be used as an indicator of the resonant activation. 
One can rely on the median of the first passage time density which is depicted in the middle panel of Fig.~\ref{fig:a19-a11}. The median location is not very sensitive to the changes in the switching rate $\gamma$ of the dichotomous process. It clearly indicates different character of slow (small $\gamma$) and high (large $\gamma$) switching rate asymptotics. Nevertheless, median location confirm presence of the resonant action for $\alpha \lessapprox 2$ only.
Better sensitivity display the inter-quantile width ($q_{0.9}-q_{0.1}$) of the first passage time density, which provides the information about the width of the interval containing $80\%$ of first passage times. The inter-quantile width, in accordance with the mean first passage time, clearly demonstrates non-monotonous dependence on the switching rate $\gamma$.
The minimal value of the inter-quantile width corroborate that escape kinetics is the most optimal, i.e. the first passage time density has the minimal width.

Figure~\ref{fig:a09-a07} demonstrates quantifiers of the resonant activation for $\alpha<1$. The left panel of Fig.~\ref{fig:a09-a07} displays the mean first passage time while the right presents the inter-quantile width of the first passage time density. Various rows correspond to various values of the stability index $\alpha$ ($\alpha\in\{0.9,9.7,05\}$) (from top to bottom). For $\alpha<1$, both the mean first passage time and the inter-quantile width indicate disappearance of the resonant activation in the system at hand.

\begin{figure}[!ht]
\begin{tabular}{cc}

\includegraphics[angle=0,width=0.95\columnwidth]{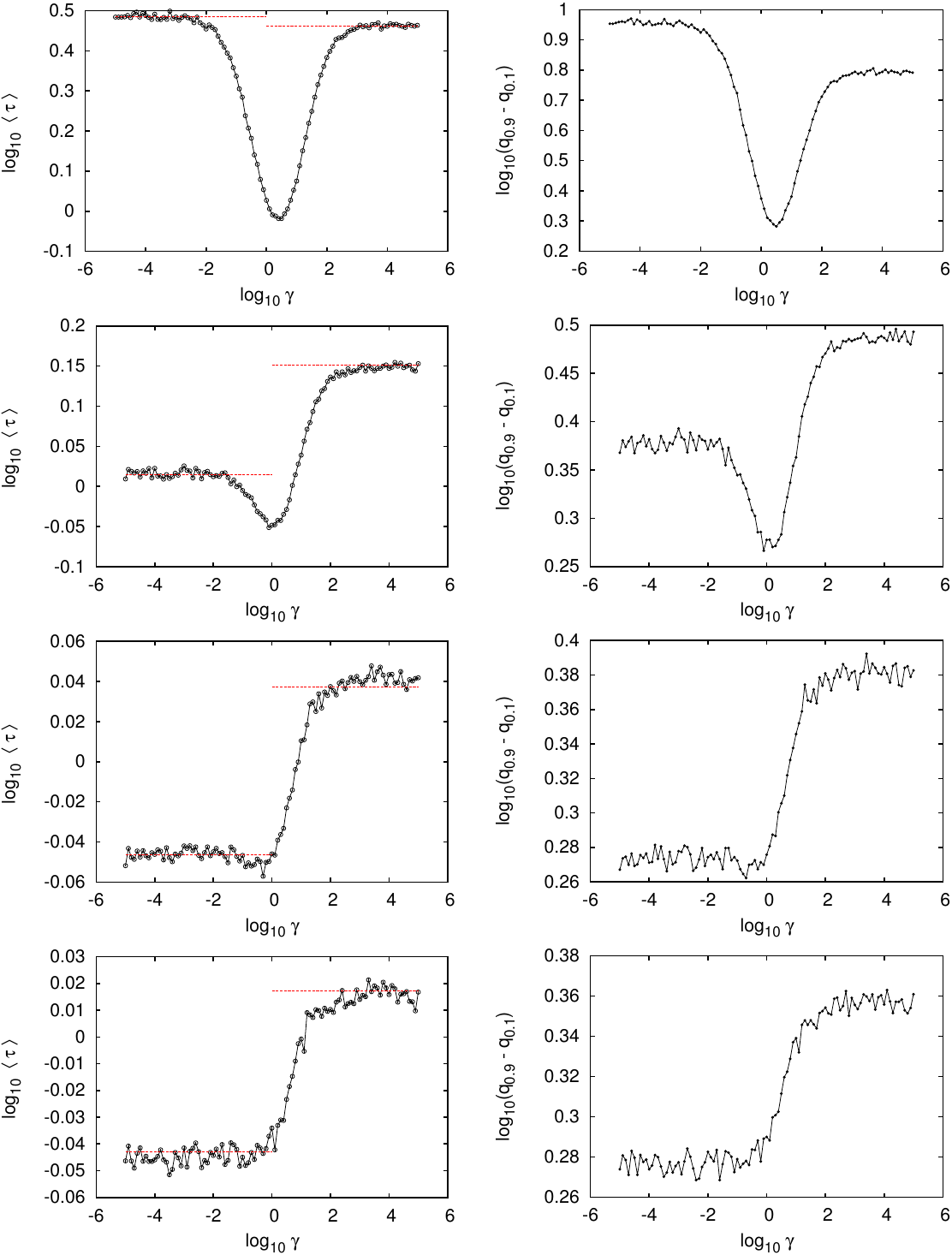} \\

 \end{tabular}
 \caption{Performance measures of the 3D resonant activation: mean first passage time $\langle \tau \rangle$ (left column) and inter-quantile width of the first passage time distribution $q_{0.9}-q_{0.1}$ (right column). Various rows correspond to decreasing values of the stability index $\alpha$ $\alpha \in \{1.9,1,5,1.1,0.9\}$ (from top to bottom).
 The potential dichotomously fluctuates between $V_\pm(\boldsymbol{r})=H_\pm r=H_\pm\sqrt{x^2+y^2+z^2}$ configurations with $H_+=8$ and $H_-=0$. The system is driven by the tri-variate $\alpha$ stable noises.
 Red dashed lines in the left column present low switching and high switching rate asymptotics.
 Error bars in the left panel represent standard deviation of the mean.
}
\label{fig:3d}
\end{figure}

The dichotomous modulation of the potential $V_\pm(r)$ results in two limiting asymptotics. On the one hand, for a very low switching rate $\gamma$ the changes in the potential are so slow that a random walker practically does not see the potential modulation. On the other hand, for a high switching rate $\gamma$ changes in the potential are so rapid that the effective height of the potential barrier is equal to the average height of the potential barrier, i.e. the potential is equal to the average potential $V(\boldsymbol{r})=\frac{1}{2}\left[ V_-(\boldsymbol{r}) + V_+(\boldsymbol{r}) \right]$ and the force acting on a particle is equal to the average force.
These two asymptotic limits are reconstructed in the mean first passage time. For $\gamma \to 0$ the mean first passage time is equal to the average of mean first passage times over both barrier configurations, because fluctuations of the potential barrier are slower that average escape time and the potential essentially stays constant during escape. Analogously, for $\gamma\to\infty$, the mean first passage time is equal to the mean first passage time over an average potential barrier, because fluctuations of the potential barrier are so fast that a random walker experiences the average potential. These asymptotic properties are recorded regardless of the presence of the resonant activation, see Fig.~\ref{fig:a19-a11} and~\ref{fig:a09-a07} which show limiting behavior of the mean first passage time (dashed red lines).

\begin{figure}[!ht]
\begin{tabular}{cc}
\includegraphics[angle=0,width=0.95\columnwidth]{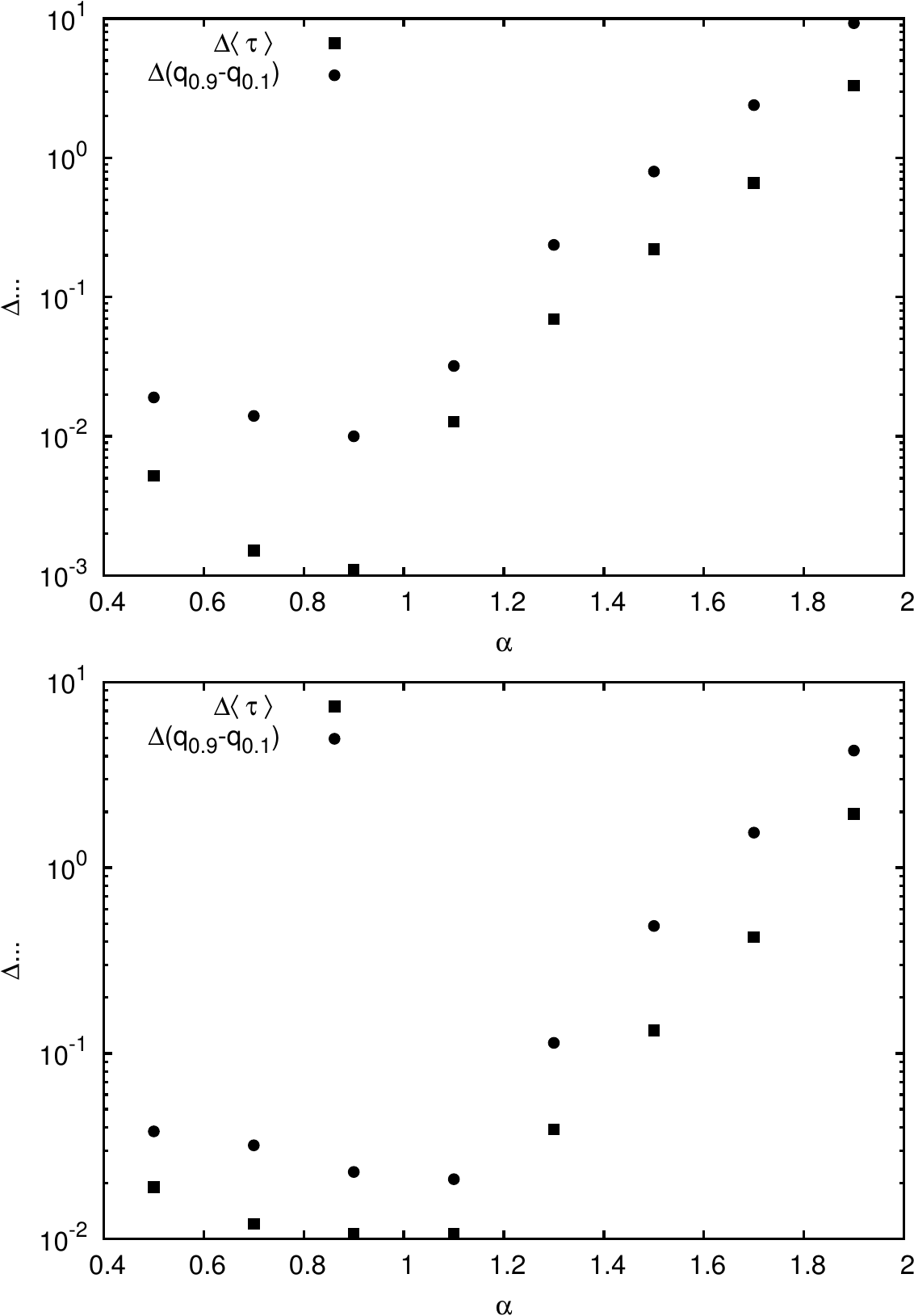} \\
 \end{tabular}
 \caption{
 Strength of resonant activation as measured by the relative depth of: mean first passage time $\langle \tau \rangle$  and inter-quantile width $q_{0.9}-q_{0.1}$ for 2D (top panel) and 3D (bottom panel) setups.
}
\label{fig:extdata-3d}
\end{figure}
 
In addition to 2D systems 3D models were considered. In particular, the tri-variate $\alpha$-stable noise driven escape from the sphere over a fluctuating potential barrier $V_\pm(\boldsymbol{r})=H_\pm r = H_\pm\sqrt{x^2+y^2+z^2}$ has been studied. Fig.~\ref{fig:3d} presents the mean first passage time (left panel) and the inter-quantile width (right panel).
In the same manner like in the 2D case, the whole surface of the sphere is absorbing and a random walker starts its diffusive motion in the sphere center, i.e. $\boldsymbol{r}(0)=\boldsymbol{0}$.
The sphere radius, $R$, is $R=1$.
In the 3D case, analogously like in the lower number of dimensions, resonant activation disappears with the decreasing value of the stability index $\alpha$. This can be confirmed both by the examination of the mean first passage time and the inter-quantile width, see Fig.~\ref{fig:3d}.
Moreover, increase in the system dimensionality further facilitate escape kinetics in comparison to 2D systems.

The strength of the resonant activation can be characterized by the deviation of the minimal value of a given quantifier from the lowest asymptotics of that quantifier, i.e.
\begin{equation}
 \Delta(u)=\min\{u(\gamma=-\infty),u(\gamma=\infty)\}-\min(u),
\end{equation}
where $u$ could be the mean first passage time, $\langle \tau \rangle$, or the inter-quantile width, $q_{0.9}-q_{0.1}$. Such a characteristics quantifies the separation of the minimum (if it exists) of a given characteristics from its lower asymptotics, i.e. from minimum of its value at low ($\gamma \to 0$) and high ($\gamma\to\infty$) switching rate $\gamma$. 
Fig.~\ref{fig:extdata-3d} corroborate that with decreasing of the stability index $\alpha$ minima of resonant activation quantifiers become less separated from lower asymptotics and finally the effect of resonant activation disappears.
The disappearance of the resonant activation for $\alpha<1$ is manifested by small, fluctuating deviations of the minimal value of resonant activation measures.

\section{Summary and Conclusions\label{sec:summary}}

Resonant activation is one of typical effects demonstrating constructive role of noise. The phenomenon of resonant activation is not only observed in 1D systems but also in higher dimensional systems perturbed by multi-variate $\alpha$-stable noises. Detailed studies of noise facilitated escape kinetics from a disk confirmed presence of resonant activation for a 2D analog of the classical resonant activation setup, i.e. escape from a bounded domain subject to the noise and external dichotomous force. Presence of the optimal escape kinetics can be corroborated not only by the mean first passage time analysis but also by inspection of quantiles of the first passage time density. With the decreasing value of the stability index $\alpha$ the effect of resonant activation weakens. Finally, for small values of the stability index, resonant activation disappears.
The same type of behavior is observed for the escape from the sphere over a fluctuating potential barrier.

Here, the resonant activation was studied for the simplest 2D and 3D setups possible, i.e. escape from a disk or a sphere with absorbing boundaries driven by bi-variate or tri-variate $\alpha$-stable noises with uniform spectral measures. Consequently, it is possible to choose other spectral measures $\Gamma(\cdot)$, e.g. non-uniform or discrete, which affect properties of noise induced jumps. Finally, it is also possible to consider other shapes of 2D domains, different boundary conditions or potentials.


%


\begin{acknowledgments}
Computer simulations have been performed at the Academic
Computer Center Cyfronet, Akademia G\'orniczo-Hutnicza (Krak\'ow, Poland) under CPU grant
MNiSW/Zeus\_lokalnie/UJ/052/2012.

\end{acknowledgments}


\def\url#1{}

\end{document}